\documentclass[aps,prd,nofootinbib,twocolumn,superscriptaddress,preprintnumbers,balancelastpage,longbibliography,nobibnotes]{revtex4-2}

\usepackage{graphicx,epsfig,psfrag,bm,amssymb,adjustbox}
\usepackage{dcolumn}
\usepackage{bm}

\usepackage{braket}
\usepackage{mathrsfs,amsfonts,hepunits,color}
\usepackage{hyperref}

\usepackage{physics}

  \usepackage{tabularx,booktabs}

\usepackage{xcolor}
\newcommand{\cw}[1]{\textcolor{black}{#1}}

\newcommand{\rev}[1]{\textcolor{black}{#1}}
\newcommand{\revv}[1]{\textcolor{black}{#1}}

\begin{document}

%Title of paper
\title{Millikelvin measurements of permittivity and loss tangent of lithium niobate}

\author{Silvia Zorzetti}
\email[email: ]{zorzetti@fnal.gov}
\affiliation{Fermi National Accelerator Laboratory, Batavia, IL 60510, USA}

\author{Changqing Wang}
\affiliation{Fermi National Accelerator Laboratory, Batavia, IL 60510, USA}

\author{Ivan Gonin}
\affiliation{Fermi National Accelerator Laboratory, Batavia, IL 60510, USA}

\author{Sergey Kazakov}
\affiliation{Fermi National Accelerator Laboratory, Batavia, IL 60510, USA}

\author{Timergali Khabiboulline }
\affiliation{Fermi National Accelerator Laboratory, Batavia, IL 60510, USA}

\author{Alexander Romanenko}
\affiliation{Fermi National Accelerator Laboratory, Batavia, IL 60510, USA}

\author{Vyacheslav P Yakovlev}
\affiliation{Fermi National Accelerator Laboratory, Batavia, IL 60510, USA}

\author{Anna Grassellino}
\affiliation{Fermi National Accelerator Laboratory, Batavia, IL 60510, USA}

\begin{abstract} 
Lithium Niobate is an electro-optic material with many applications in microwave signal processing, communication, quantum sensing, and quantum computing. In this letter, we present findings on evaluating the complex electromagnetic permittivity of lithium niobate at millikelvin temperatures. Measurements are carried out using a resonant-type method with a superconducting radio-frequency (SRF) cavity operating at 7~GHz and designed to characterize anisotropic dielectrics. The relative permittivity tensor and loss tangent are measured at 50~mK with unprecedented accuracy.

\end{abstract}
%\pacs{}

\maketitle

\makeatletter
\def\l@subsubsection#1#2{}
\makeatother

\textit{Introduction.} Lithium Niobate (LN: LiNb$\text O_\text3$) is a nonlinear ferroelectric material with \cw{rich} optical, acoustic, and piezoelectric properties \cite{weis1985lithium}. Because of the large electro-optic coefficient (r33 = 31~pm$/$V at 9~GHz), it is of great interest in \cw{various} applications spanning from communication, microwave-optical transduction, biomedical devices to sensing~\cite{han2021microwave, lauk2020perspectives, wang, levi, milazzo, lauk}.
As a \cw{basic element of an} electro-optic modulator (EOM), LN modulates an incoming optical career signal by the amplitude of an applied voltage due to the Pockels effect. Optical cavities made of LN can perform with a high quality (Q) factor in the range of $10^6 - 10^8$~\cite{zhang2019broadband}. This is possible through high-quality chemical and mechanical polishing of the surfaces. Such optical cavities can operate in whispering gallery modes, \cw{where a three-wave mixing process can occur between two optical modes and a microwave mode} under the phase-matching condition~\cite{tsang2010cavity}. Single-photon level measurements at millikelvin temperatures have demonstrated the ability of LN to maintain good electro-optic modulation properties in this regime~\cite{fink, lauk, sahu2023entangling}. \cw{Furthermore, the piezoelectric properties of LN allow the coupling between the electric signal and the mechanical degree of freedom, leading to intriguing phenomena in piezo-optomechanics~\cite{mirhosseini2020superconducting}.
Therefore, LN can be exploited to design physical devices to leverage the interaction between microwave, mechanical, and optical fields, which operate at intrinsically disparate frequencies.}  

Among the few materials exhibiting the Pockels effect and \cw{piezoelectric property}, LN is one of the most employed in emerging applications in quantum computing and sensing for the coherent transduction of single photons~\cite{fink,mirhosseini2020superconducting} and for enhancing the sensing capabilities of quantum sensors. These applications \cw{rely on bulk LN in three-dimensional (3D) hybrid systems or thin-film LN compatible with integrated two-dimensional (2D) devices}. While several studies exist on the properties of LN crystals at room or higher temperatures~\cite{jazbinvsek2002material,cena2016structural, ohmachi1967dielectric, abrahams1989properties}, there is a need to gain more knowledge of the complex electromagnetic permittivity at cryogenic and millikelvin temperatures. Electromagnetic characterization of such dielectric crystals is necessary to understand their behavior under applied electromagnetic fields and, therefore, to design and devise many quantum computing and sensing applications and evaluate performance accurately. Goryachev et al. have determined the loss tangent of lithium niobate in the order $10^{-5}$ at the single photon level~\cite{Tobar}. \cw{Besides, Wollack et al. studied the loss mechanisms of thin-film LN, including the resonant two-level-system (TLS) decay, the off-resonant TLS relaxation damping, and the temperature-independent mechanical loss at cryogenic temperatures~\cite{wollack2021loss}.} In this letter, we present the measurement results of the complex dielectric properties of \cw{bulk LN} crystals \cw{for 3D electro-optic applications} at the quantum threshold with reduced uncertainty.  

\begin{figure}[!htb]
    \centering
    \includegraphics[width=0.45\textwidth]{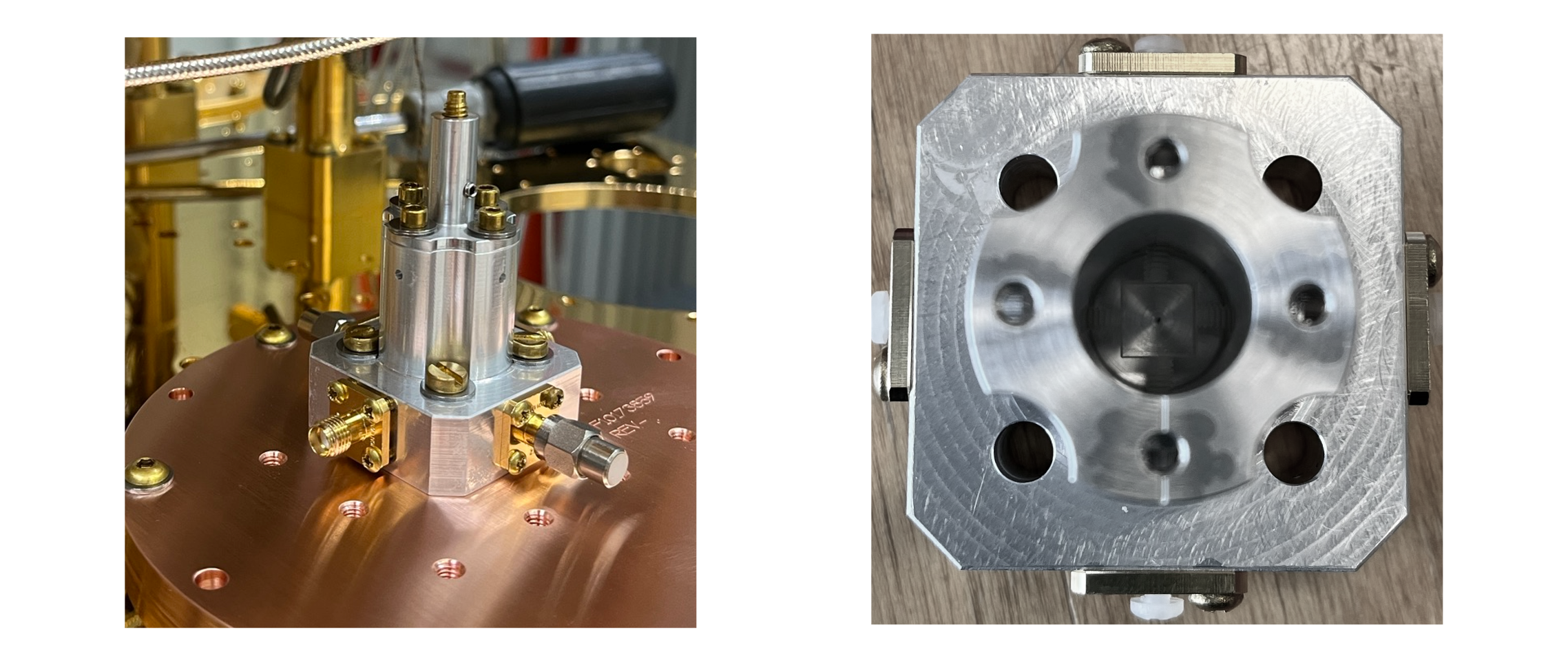}
    \caption{\textit{Left:} Sample holder cavity in a dilution refrigerator (DR). \textit{Right:} Top view of the cavity with alignment features during installation of the LN crystal.}
    \label{fig:pict}
\end{figure}

\textit{Methods.} Superconducting radio-frequency (SRF) cavities are developed in high energy physics (HEP) to accelerate particle beams up to the speed of light. Such cavities can also achieve very high quality factors ($Q>10^{10}$) at millikelvin temperatures~\cite{romanenko, romanenko2017understanding}. Because of the low microwave loss, SRF cavities are valuable tools for characterizing dielectric materials with high accuracy~\cite{checchin, zorzetti2022methods,mcrae2020materials}. \cw{Here,} an SRF cavity made of aluminum is designed and manufactured to study the loss tangent and the permittivity \cw{of} bulk samples of lithium niobate or other ferroelectric materials  (Fig.~\ref{fig:pict}). The cylinder-shaped cavity hosts a cubic sample\rev{\footnote{\rev{The sample is made of pure single-crystal LN and has a dimension of $6\times6\times6$ mm. The z-cut surfaces are polished with surface quality of 20/10 scr/dig, flatness of  $\lambda$/8, and parallelism $<20$ arcsec. }}} of a z-cut LN. Alignment features are used to place the sample at the center of the cavity. 

\begin{figure}[!htb]
    \centering
    \includegraphics[width=0.45\textwidth]{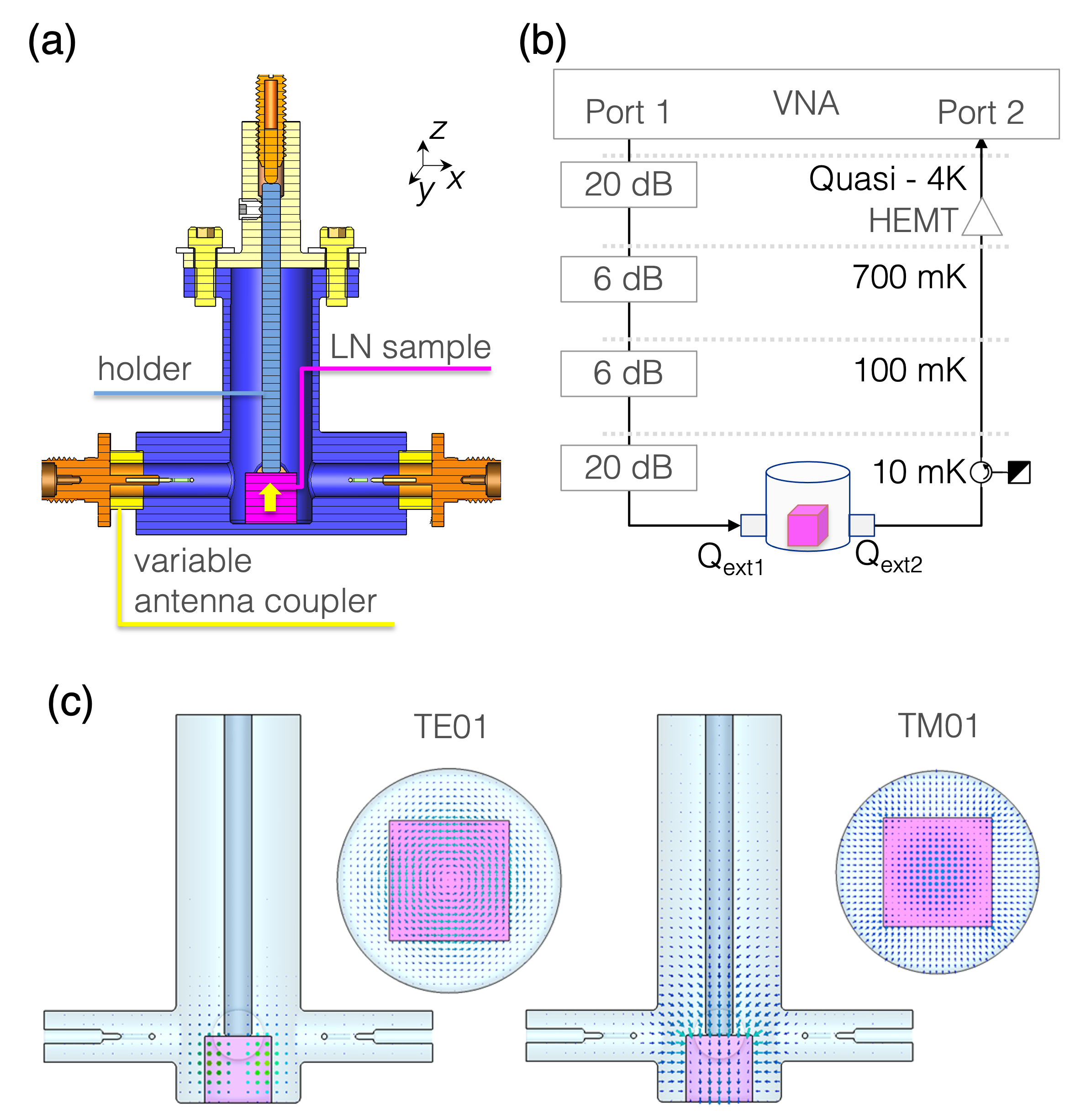}
    \caption{\textbf{(a)} Model of the cavity with the crystal embedded. The z-cut face of the LN crystal (\cw{perpendicular to} the yellow arrow) is aligned with the $z-$axis of the SRF cavity. \textbf{(b)} Schematic view of the experimental setup in the dilution refrigerator. \rev{\textbf{(c)} Electric field distributions of the TE01 and TM01 modes along the $z-$axis and the $xy-$plane.}}
    \label{fig:BD}
\end{figure}

The cavity is equipped with four lateral variable antenna couplers (Fig~\ref{fig:BD}(a)). \rev{Loop antenna probes are used to couple to modes with different polarizations}. %This system is designed to improve the thermal conductivity between the sample and the cavity for better thermalization. 
The bulk sample \rev{is not glued, it }is held in position through a spring-loaded sapphire rod \rev{that compensates for thermal shrinkage during cooldown, allowing to keep a constant force on the sample from room temperature through cryogenic temperature, and} improves the thermal conductivity between the sample and the cavity. Sapphire is a dielectric material with a negligible loss tangent compared to the sample under test. Specifically, the loss tangent of sapphire is $\sim10^{-8}$ at low temperatures~\cite{krupka1999complex, creedon2011high}. The diameter of the holder is selected in such a way that there is no radiation trough the holder pipe at the operating frequencies. The cavity is installed in a dilution refrigerator (DR) and cooled down to millikelvin. The schematic view of the experiment in the DR is shown in Fig.~\ref{fig:BD}(b). As the environmental conditions, such as temperature or microwave power, vary, the changes in the quality factor and the frequency shifts are monitored to evaluate \cw{the dielectric permittivity} and the crystal's loss tangent.

Through a VNA (Vector Network Analyzer), the loaded quality factor \cw{of the cavity with the sample enclosed} ($Q_L$) is measured. $1/Q_L$ is the sum of several components:

\begin{equation}
    \frac{1}{Q_L}=\frac{1}{Q_0}+\frac{1}{Q_d}+\frac{1}{Q_{ext1}}+\frac{1}{Q_{ext2}}\cw{+\frac{1}{Q_{ext3}}+\frac{1}{Q_{ext4}}}.\label{Q_L3}
\end{equation}
$Q_0$ is the intrinsic quality factor of the cavity, which takes into account the microwave loss on the cavity walls. \cw{$Q_{ext1,2,3,4}$} are the external quality factors of the couplers, which are calibrated at room temperature. \cw{In our experiments, the ports 3 and 4 are decoupled from the cavity such that $\frac{1}{Q_{ext3,4}}\rightarrow0$.} The \cw{other two} couplers are tuned in such a way that the cavity is undercoupled. $Q_d$ is the dielectric quality factor, i.e. the energy stored in the cavity ($W$) over the power dissipated in the dielectric \cw{sample} ($P_d$), multiplied by the resonant frequency ($\omega$):
\begin{equation}
    \label{Q_L2}
    Q_d=\frac{\omega W}{P_d}=\frac{1}{p\times\tan\delta},
\end{equation}
where, $p$ is the filling factor \cw{of the dielectric material}, i.e. \rev{the energy stored in the sample divided by the total stored energy in the whole cavity. }
%the fraction of the microwave loss in the dielectric sample compared to the total losses. 
The filling factor also depends on the dielectric permittivity of the crystal ($\varepsilon_r$), and it is different for each of the modes excited in the cavity. 

Once the dilution refrigerator reaches the superconducting critical temperature of aluminum ($\text{T}_\text{C}$ = 1.2~K), the microwave loss on the cavity's walls are negligible compared to the dielectric loss in the LN sample ($Q_0>>Q_d$). The unloaded quality factor ($Q_{UL}$) is therefore calculated as:
\begin{equation}
    \frac{1}{Q_{UL}}=\frac{1}{Q_L}-\frac{1}{Q_{ext1}}-\frac{1}{Q_{ext2}}\simeq\frac{1}{Q_d}.\label{Q_UL}
\end{equation}

The complex permittivity in isotropic materials can be written as $\varepsilon=\varepsilon_0\varepsilon_r=\varepsilon_0{\varepsilon'}_r(1-j\tan\delta)$. The relative permittivity in anisotropic non-centrosymmetric crystals such as LN is instead a tensor:
\begin{equation}
\varepsilon_r=\begin{pmatrix}
\varepsilon_\perp & 0 & 0\\
0 & \varepsilon_\perp & 0\\
0 & 0 & \varepsilon_\parallel
\end{pmatrix},\label{eps_tensor}
\end{equation}
where, $\perp$ and $\parallel$ refer to the  components \cw{perpendicular and parallel to the principal axis of the crystal}, respectively \cite{li2022complex}. %\cw{When there are small changes in the dielectric tensor components, the}
The frequency shifts in resonant modes are proportional to the \cw{changes in the} dielectric constant:
\begin{equation}
    \Delta f = \cw{\frac{df}{d\varepsilon_{\perp}} \Delta \varepsilon_{\perp}+\frac{df}{d\varepsilon_{\parallel}} \Delta \varepsilon_{\parallel}},
\end{equation}
where \cw{$\frac{df}{d\varepsilon_{\perp,\parallel}}$ are} calibration constants derived from electromagnetic (EM) simulations for each of the microwave modes. With the dielectric crystal occupying most of the electromagnetic volume, losses are mainly concentrated in the crystal. Therefore, with $Q_L~\sim~10^5$, the measurement's resolution in the evaluation of the dielectric constant is up to $\delta\varepsilon=0.01$.

\cw{The z-cut face is perpendicular to} the $z-$axis of the cavity \cw{which is the principal axis of the crystal} (as in Fig.~\ref{fig:BD}(a)). Among the various modes excited in the cavity, we monitor the TM01 and TE01 modes. \rev{The electric field distributions along the $z-$axis and the $xy-$plane are shown in Fig.~\ref{fig:BD}(c)}. While the electric field of the TM01 mode is parallel to the \cw{$z-$axis}, the electric field of the TE01 mode has components only in the $xy-$plane. For each mode, the dielectric loss tangent is evaluated as:
 \begin{equation}
     \frac{1}{Q_d}=p_\perp\tan\delta_\perp+p_\parallel\tan\delta_\parallel.
 \end{equation}
\cw{For TM and TE modes,} the electric field is aligned to the $\parallel$ or $\perp$ planes, respectively \cite{li2022complex}. \cw{Hence, we obtain}:
 \begin{equation}
\begin{split}
 &\frac{1}{{Q_d}^{(TE)}}=p_\perp\tan\delta_\perp,\\ 
 &\frac{1}{{Q_d}^{(TM)}}=p_\parallel\tan\delta_\parallel.
\end{split}\label{tg_tensor}
\end{equation}

%For this specific case, the cavity is mostly filled with the dielectric sample of lithium niobate. In such hybrid cavities, with high filling factor, the quality factor is therefore mostly driven by the microwave losses in the crystal ($Q_0\simeq 1/\tan(\delta)$). For this study, we use two cavities, one made of niobium (Nb) and other of aluminum (Al). The z-cut sample is hosted in the Al cavity and the z-cut is hosted in the Nb one. One, it is convenient having two cavities available to measure two samples in parallel, it is also a chance to compare the performance of the two superconducting metals, in cases in which the filling factor is high. 

\begin{figure*}[htb]
    \centering
    \centerline{\includegraphics[width=\textwidth]{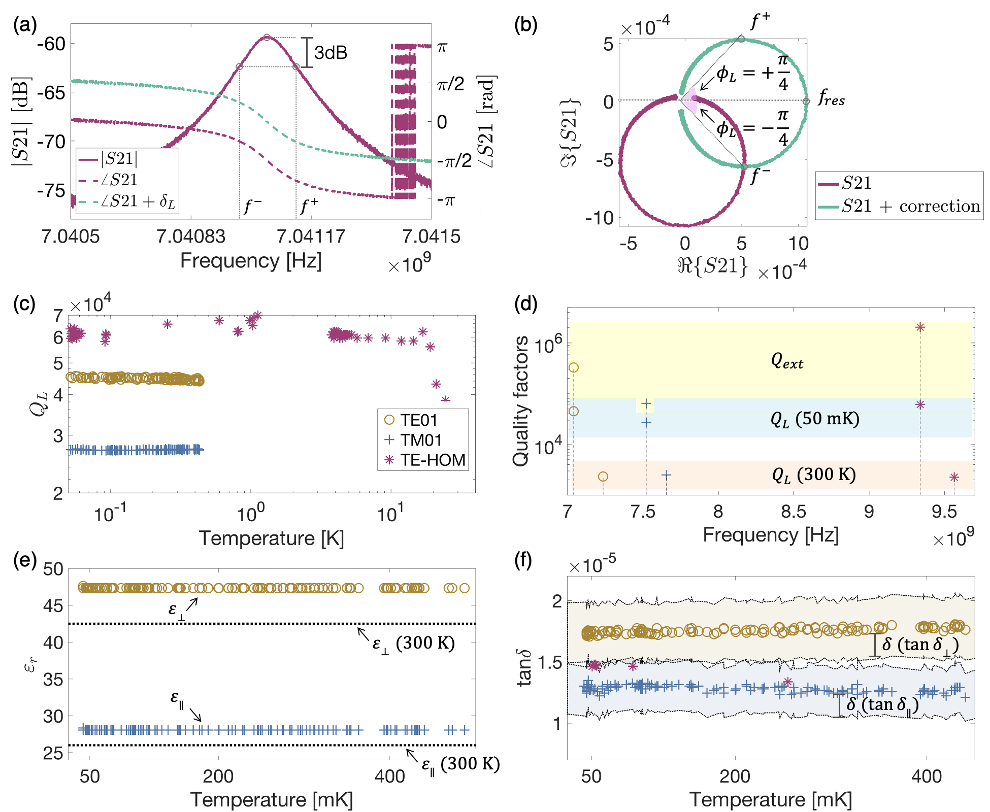}}
        \caption{\rev{\textbf{(a)} Magnitude and phase components of the S21 parameter. A phase correction is applied to cancel the detuning angle. \textbf{(b)} In-phase and quadrature components of the S21 parameter before and after the phase correction. Relevant RF parameters for the evaluation of the loaded quality factor are noted.} \textbf{(c)} Loaded quality factor vs. temperature. \textbf{(d)} External quality and loaded quality factors at room temperature and at cryogenic temperatures, for each mode. \textbf{(e)} Measured permittivity vs. temperature. \textbf{(f)} Measured loss tangent vs. temperature. The total uncertainty is represented by the shaded area.}\label{fig:QL_Q0_Qe_tg}

\end{figure*}

\textit{Measurements.} Transmission parameters ($S21$) of TE and TM modes \cw{are measured using a VNA}. The loaded quality factor is computed as $Q_L=\frac{f_{res}}{BW}$, \rev{where $BW$ is the 3dB bandwidth. The 3dB bandwidth is estimated by applying a phase correction to rotate the in-phase and quadrature circle, canceling the detuning angle ($\delta_L$). Then, the bandwidth is computed as the difference between $f^+$ and $f^-$: the locus point in which the absolute value of the imaginary part of S21 is at a maximum, corresponding to two vectors with arguments of $\phi_L=\pm\pi/4$ \cite{Kajfez, Caspers}, as shown in \ref{fig:QL_Q0_Qe_tg}(a-b).}

%which are the points at $\phi_L=\pm\pi/4$, as well as the points in which the absolute value of the imaginary part is at its maximum \cite{Kajfez, Caspers}, as shown in \ref{fig:QL_Q0_Qe_tg}(a-b).}

Measurements of external and loaded quality factors are shown in Fig.~\ref{fig:QL_Q0_Qe_tg}(c-d). As $Q_0>>Q_d$\rev{\footnote{\rev{It is noted that the surface of the aluminum cavity was treated with buffered chemical polishing (BCP), removing \SI{10}{\micro\meter} on the surface to reduce the losses on the superconducting cavity's walls}. }}the quality factor of the dielectric sample is calculated as in Eq.~(\ref{Q_UL}). Frequency shifts are monitored, and changes in the electrical permittivity are referenced to the real part of the electrical permittivity at room temperature $\varepsilon_r~=~(42.5,~42.5,~26)$~\cite{ohmachi1967dielectric, abrahams1989properties}. 

The filling factors ($p_\perp$ and $p_\parallel$) are computed through EM simulations using the values of the measured $\varepsilon_r$ at 50~mK. To complete the analysis, the loss tangent is calculated from Eq.~(\ref{tg_tensor}). The measurements of the relative permittivity and the loss tangent at $\text T<$500~mK, which are the main findings of this paper, are displayed in Fig.~\ref{fig:QL_Q0_Qe_tg}(e-f).

We confirm that the loss tangent of LN is in the order of $10^{-5}$ at microwave frequencies, and notably, we narrow the uncertainty range. In particular, we obtain $\tan\delta_\perp=$~$1.73\times 10^{-5}$ and $\tan\delta_\parallel=$~$1.28\times 10^{-5}$ at 50~mK. Main design parameters and measurements results are summarized in Table~\ref{par_table}. 

Furthermore, we acquire few data points of a TE-like higher-order mode (HOM). The electrical field of this resonant mode propagates mainly on the \textit{xy}-plane, but there are also components on the $z-$axis at a frequency of 9.5~GHz. The resulting loss tangent falls in between $\tan\delta_\perp$ and $\tan\delta_\parallel$ \cw{(Fig.~\ref{fig:QL_Q0_Qe_tg}(c)} and Fig.~\ref{fig:QL_Q0_Qe_tg}(f)).

\begin{table}[h]
	\centering
	\caption{Main design parameters and measurements results.}\label{par_table}
	\begin{tabularx}{0.45\textwidth}{X| X X}
	\hline\hline
		Parameter&Value&Unit\\\hline
            LN sample   & $6\times6\times6$ & mm\\
            Cavity ($D\times H$)  & 11.2 $\times$ 34.5 & mm   \\
            \hline
            $f^{(TE01)}$ & 7.2 & GHz \\
            $f^{(TM01)}$ & 7.6 & GHz\\
            
        \hline
            $df/d\varepsilon_\perp$ & -40.603&MHz\\ 
            $df/d\varepsilon_\parallel$ &-64.637 &MHz \\
            \hline
            $\varepsilon_\perp$ & 47 &\\
            $\varepsilon_\parallel$ & 28 &\\
            \hline
            
            $p_\perp$ (50~mK) & 92.3 & $\%$ \\
            $p_\parallel$  (50~mK) &  46.3 & $\%$\\
            \hline

            $\tan\delta_\perp$ (50~mK) &$1.73\times 10^{-5}$ & \\
            $\tan\delta_\parallel$ (50~mK) & $1.28\times 10^{-5}$& \\

		\hline\hline
	\end{tabularx}	
\end{table}

\textit{Error Analysis.} The primary sources of error in the presented analysis are identified in the external quality factors and in the filling factors. \rev{To evaluate the uncertainty in the estimation of $Q_{ext}$, we compared measured and simulated values.} \rev{We simulated $Q_{ext}$ based on the knowledge of the antenna position and material properties.} 
Then, we measured $Q_{ext}$ through reflection on the 4-port cavity. These measurements show that additional radiations might be present, particularly for the TM01 mode, which propagates on the $z-$axis. Therefore, simulated and measured values are considered as the two extreme cases. We calculate $Q_d$ from Eq.~(\ref{Q_UL}) for both cases, and we found the uncertainty of $Q_d$ ($\delta(Q_d)$) for the TM and TE modes.

%We find a disagreement between the measured values and the EM simulation. This might be due to additional radiations, in particular for the TM01 mode, that propagates on the z-axis. 
%\rev{Measured values of $Q_{ext}$ are compared to EM simulations. These are considered}
%Therefore, we assume EM simulations and measurements 
%We consider the measurement uncertainty of $Q_L$ to be negligible compared to the uncertainty in the estimation of $Q_{ext}$. (as in Eq.~(\ref{Q_L}).% 
%For simplicity, we assume $Q_{ext1}=Q_{ext2}$.
%\begin{equation}
%\begin{split}
% \frac{\delta(Q_0)}{Q_0}=&\\
% &\sqrt{\left(\frac{\delta(Q_{L})}{Q_L}\right)^2+2\left(\frac{\delta(Q_{ext})}{Q_ext}\right)^2+\left(\frac{\delta(Q_{ext}-Q_L)}{Q_{ext}-Q_L}   \right)}\simeq\\
% &\sqrt{3}\left( \frac{\delta(Q_{ext})}{Q_{ext}}\right)
% \end{split}
%\end{equation}
%\begin{equation}
% \frac{\delta(Q_d)}{Q_d}=\sqrt{2}\left(\frac{\delta(Q_{ext})}{Q_{ext}}\right).
%\end{equation}
The uncertainty in the measurement of the loss tangent is derived as a combination of the estimation of the dielectric loss and the filling factor:

\begin{equation}
\frac{\delta(\tan\delta)}{\tan\delta}=\sqrt{\left( \frac{\delta(p)}{p} \right)^2 + \left( \frac{\delta(Q_d)}{Q_d} \right)^2}.
\end{equation}

The uncertainty in the filling factor (\cw{$\delta(p)$}) is also estimated through EM simulations. We consider an extreme case, in which the crystal is misaligned in the cavity with an offset of 0.5~mm. The total uncertainty is represented by the shaded area in Fig.~\ref{fig:QL_Q0_Qe_tg}(f). The results of the error analysis are listed in Table~\ref{table_unc}.

\begin{table}[ht]
\centering
\caption{Estimated uncertainties.}\label{table_unc}
\begin{tabularx}{0.45\textwidth}{X| X X}
\hline\hline
Variable (V) & \textbf{$\delta(V)/V$}& \textbf{ $\delta(V)$ }\\
    \hline
    
 $Q_d$ (TE01) & 0.14 & $\sim8 \times 10^3$ \\  
  $Q_d$ (TM01) & 0.15 & $\sim2.6 \times 10^4$ \\  

 \hline
  $p_\perp$ & 0.002 & 0.19\\
 $\tan\delta_\perp$ & 0.14 & $\sim 0.24 \times 10^{-5}$\\
  \hline
  $p_\parallel$ & 0.049 & 2.22\\
 $\tan\delta_\parallel$ & 0.16 & $\sim 0.21 \times 10^{-5}$\\

 \hline\hline
\end{tabularx}
\end{table}

\begin{figure}[!htb]
    \centering
    \centerline{\includegraphics[width=0.45\textwidth]{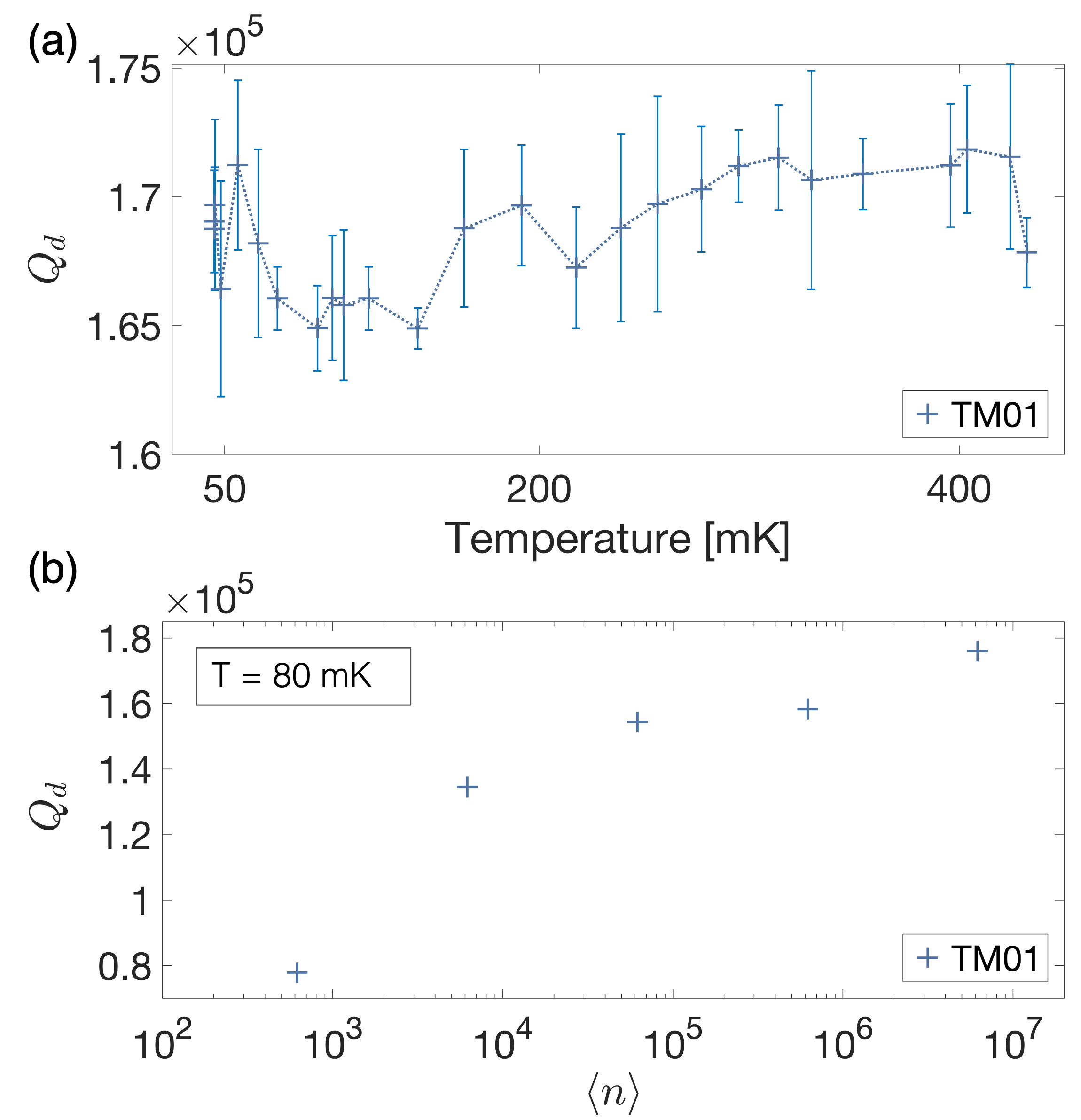}}
    \caption{\textbf{(a)} Losses in the LN crystal as a function of temperature. The error bars represent the standard deviation of the data acquired and analyzed. \rev{The input power is $P_{in}=$\SI{-52}{dBm}.} \textbf{(b)} Losses in the LN crystal as a function of the average number of photons. $P_{in}$ spans from \SI{-92}{} to \SI{-52}{dBm}.}
    \label{fig:Qd}
\end{figure}

\textit{\cw{Loss Channels}}. \cw{The microwave loss in a piezoelectric dielectric sample may include the TLS (two-level-system) loss, the quasi-particle loss, the piezoelectric loss, and others \cite{scigliuzzo2020phononic}. To investigate the loss mechanisms in the LN sample, we analyze the dielectric quality factor of the TM mode versus the temperature  (Fig.~\ref{fig:Qd}(a)). The weak temperature dependence of $Q_d$ indicates that the quasi-particle loss is negligible in the overall cavity~\cite{scigliuzzo2020phononic}. We then} measure the quality factor at different input power levels. The average number of photons stored in the cavity $\left< n\right>$ is determined through the transmitted power: $P_tQ_{ext2}~=~\cw{{\hbar\omega^2\left< n\right>}}$, with $P_t=P_{in}-P_r-P_{loss}$, where $P_{in}$ is the input power, $P_r$ is the reflected power at Port~1, $P_t$ is the power transmitted to Port~2, and $P_{loss}$ is the power dissipated in the cavity~\cite{melnychuk2014error, romanenko2017understanding}. Fig.~\ref{fig:Qd}(b) displays $Q_d$ versus the average number of photons stored in the cavity $\left<n\right>$ at a fixed temperature of 80~mK. We observe that the dielectric quality factor \cw{increases} with the number of photons. This is clear evidence of \cw{ the presence of TLS as a source of dissipation, because the TLS is saturated at high RF power~\cite{wollack2021loss,scigliuzzo2020phononic}}. Moreover, a temperature-independent piezoelectric loss and other types of unknown losses may exist. Further studies may be carried out to disentangle TLS and non-TLS mechanisms by precisely scanning the temperature and amplitude of the microwave field in the cryostat.

\textit{Conclusions \revv{and Outlook.}} Lithium niobate and other electro-optic dielectrics are being recognized as promising materials for quantum computing and sensing. We present findings on evaluating the complex electromagnetic permittivity of lithium niobate. Using a resonant-type method, we determined the loss tangent and the relative permittivity at millikelvin temperatures with unprecedented accuracy. We measured the parallel and perpendicular components of the loss tangent to be $(1.28\pm 0.21)\times10^{-5}$ and $(1.73\pm0.24)\times 10^{-5}$, respectively. These measurements are critical to devise applications that include coherent frequency conversions of microwave-optical photons. \revv{This project motivates further investigations to understand better the loss mechanisms of lithium niobate and other noncentrosymmetric crystals and explore the low-power limit. Future studies include measurements in the broader frequency regime, which shall be enabled by tunable cavities.}

\newpage

\textit{Acknowledgments.} 
This manuscript has been authored by Fermi Research Alliance, LLC under Contract No. DE-AC02-07CH11359 with the U.S. Department of Energy, Office of Science, Office of High Energy Physics. This work is funded by the Fermilab’s Laboratory Directed Research and Development (LDRD) program.
This research used resources of the U.S. Department of Energy, Office of Science, National Quantum Information Science Research Centers, Superconducting Quantum Materials and Systems Center (SQMS) under contract number DE-AC02-07CH11359. The NQI Research Center SQMS contributed by supporting the design of SRF cavities and access to facilities. 
The authors would like to acknowledge help with the experimental setup from Roman Pilipenko and Geev Nahal Gelehpordsari, help with the mechanical design from Oleg V. Pronitchev, and help with cavity production from Edward D. Pieszchala and the Fermilab's village machine shop.

%\bibliography{bibliography}

\end{document}